\documentclass[prc,twocolumn,showkeys,showpacs,floatfix]{revtex4}

\usepackage{graphicx,amsmath}

\newcommand{\refEq}[1]{ Eq.~(\ref{#1})}

\begin{document}

\title{Isospin Fluctuations from a Thermally Equilibrated Hadron Gas} 

\author{Sen Cheng}

\author{Scott Pratt}
\affiliation{ Department of Physics and Astronomy and National Superconducting
  Cyclotron Laboratory,\\ Michigan State University, East Lansing, Michigan
  48824-1321 }

\date{\today}

\begin{abstract}
  Partition functions, multiplicity distributions, and isospin fluctuations are
  calculated for canonical ensembles in which additive quantum numbers as well
  as total isospin are strictly conserved.  When properly accounting for
  Bose-Einstein symmetrization, the multiplicity distributions of neutral pions
  in a pion gas are significantly broader as compared to the non-degenerate
  case.  Inclusion of resonances compensates for this broadening effect.
  Recursion relations are derived which allow calculation of exact results with
  modest computer time.
\end{abstract}

\pacs{02.70.Rr, 24.60.-k} 
\keywords{isospin fluctuations, disoriented chiral condensates, statistical
  nuclear physics} 

\maketitle

\section{Introduction}
\label{intro}

Motivated by the observation of large fluctuations of the ratio of neutral to
charged particles in cosmic ray events \cite{Lattes:1980wk,Borisov:1987sp},
numerous studies 
of isospin fluctuations have been undertaken during the last decade. It has
been proposed that the melting and subsequent re-condensation of the chiral
condensate could provide a dynamical means for coherent pion emission
where dozens of pions are emitted with the same
isospin. If $N$ pions are confined to a single quantum state in addition
to being in an isosinglet \cite{Horn:1971AK},
the probability of
finding $n_0$ neutral pions takes a simple form in the limit of large 
$N=n_+ + n_- + n_0$,
\begin{equation}
\label{eq:coherentisodist}
\frac{dN}{df}=\frac{1}{2\sqrt{f}},\quad f\equiv n_0/N.
\end{equation}
One can obtain the same result by considering a coherent state
\begin{equation}
|\vec{\eta}\rangle=\exp \left(\vec{\eta}\cdot\vec{\pi}\right)|0\rangle,
\end{equation}
where the pion field operators are $\pi_0=\pi_z,\ \pi_{\pm}=(\pi_x\pm i\pi_y)/
\sqrt{2}$, and the direction of $\vec{\eta}$ is averaged over all directions.
The source of the field $\vec{\eta}$ has been proposed to be the chiral
condensate which might disorient itself in a quenching scenario
\cite{Rajagopal:1993qz}. This is often referred to as disoriented chiral
condensate (DCC).

The dramatically broad isospin distribution of \refEq{eq:coherentisodist}
relies on the assumption that the emission proceeds via a single quantum state.
The inclusion of Bose-Einstein effects in the thermal emission of pions from a
non-degenerate array of states was shown to broaden the isospin distribution
\cite{Pratt:1993uy}, but not nearly as much as in \refEq{eq:coherentisodist}.
Neglecting isospin conservation in \cite{Pratt:1993uy} accounted for the
reduced broadening of the peak.  A crude accounting for isospin conservation
was suggested by considering the emission of neutral pion pairs (2/3
$\pi^+\pi^-$ and 1/3 $2\pi_0$) \cite{Pratt:1994tu}, but came far short of
considering the complete ensemble of isoscalar states in a multi-level system.
The emission of pairs through a classical isoscalar field into non-degenerate
single-particle levels, which may be considered as an oriented chiral
condensate, has been studied as well \cite{Volya:1999ut}.

The effects of exact charge conservation in canonical ensembles were also
studied in other contexts with a projection method. Strangeness and baryon
number conservation were found to enchance strangeness productions in
$\bar{p}N$ collisions, particularly for small systems \cite{Derreth:1985kk}.
The confinement of the quark-gluon plasma to color-singlets was shown to lead
to a reduction in the number of internal degrees of freedom, which could lead
to measurable finite size effects in relativistic heavy ion collisions
\cite{Elze}.

In this paper, we extend the sophistication of statistical treatments by
considering the entire ensemble of isoscalar states available in a system with
many single-particle levels.  Methods for calculating isospin distribution are
presented, which include Bose-Einstein symmetrization, the effects of
resonances and the conservation of both total isospin $I$ and its projection
$M$.  We present sample calculations to illustrate the effects mentioned above
and find that symmetrization effects are important for high quantum
degeneracies, that isospin conservation has little impact when the size of the
domain exceeds a dozen pions, and that resonances can strongly narrow the
distribution.

The next section provides a description of the recursive methods developed to
determine canonical partition functions, which include Bose-Einstein
symmetrization, isospin conservation, and resonances. Techniques for exact
calculation of isospin distributions, which are consistent with such ensembles,
are presented in Sec.~\ref{sec:multdisttheory}, results in
Sec. \ref{sec:results}, and conclusions in \ref{sec:conclusions}.

\section{Recursion Relations for Partition Functions}
\label{sec:partfunctheory}

\subsection{Non-degenerate Systems}
\label{subsec:partfunctheory_classical}

When the number of available states is much larger than the number of
particles, the probability for two or more particles occupying the same quantum
state is small. In such a non-degenerate system, quantum statistics can be
neglected.  The partition function for a canonical ensemble of $A$ particles
conserving an additive quantum number or a vector of such quantities $Q$ can be
written as a product of single-particle partition functions
\begin{equation}
  \label{eq:pf_sum_partitions}
  Z_{A,Q} = \sum_{\langle \sum \nu_k a_k = A \rangle 
    \atop \langle \sum \nu_k q_k = Q\rangle}
  \prod_{k=1}^{N} \frac{\omega_k^{\nu_k}} {\nu_k!},
\end{equation}
where $N$ is the number of particle types, $\nu_k$ is the occupation number of
particle type $k$, $q_k$ is the charge of one particle, and the particle number
$a_k$ indicates how many times a particle contributes to the main conserved
quantity $A$.  For example, if $A$ is the number of pions then $a_\rho= 2$,
since the $\rho$ meson decays predominantly into two pions.  The
single-particle partition function $ \omega_k= g_k \sum_i
\exp(-\epsilon^{(k)}_i/T) $ sums Boltzmann factors weighted by the the spin
degeneracy $g_k$ over all available single-particle levels $i$.

Summation over the immense number of partitions in \refEq{eq:pf_sum_partitions}
can be avoided by rewriting the partition function as a recursion relation
\cite{Chase:1995AK,DasGupta:1998AK,Pratt:1999ht},
\begin{equation}
  \label{eq:cl:Z}
  Z_{A,Q} =  \sum_{k=1}^N \frac{a_k \omega_k}{A}  Z_{A-a_k, Q-q_k}.
\end{equation}
If intermediate values of the partition function are stored, the computations
required for \refEq{eq:cl:Z} scale linearly in both $A$ and $N$, thus making
it possible to quickly calculate the canonical partition function numerically.

A partition function conserving total  isospin as well as additive quantum
numbers is derived by  adding a sum over all 
possible isospin configurations for a given partition, $\{\nu_k\}$, and
isospin weights $\xi\left(I,M|\{\nu_k\}\right)$ to
\refEq{eq:pf_sum_partitions},
\begin{equation}
  \label{eq:omegaaimdef}
  \Omega_{A,I,M} = 
  \sum_{ \langle \sum \nu_j a_j = A \rangle 
    \atop \langle \sum \nu_j q_j = Q \rangle}
  \sum_{\{\nu_j\}} \xi\left(I,M|\{\nu_j\}\right)
  \prod_{j=1}^{N} \frac{\omega_j^{\nu_j}} {\nu_j!}.
\end{equation}
To convert this partition function into a recursion relation insert $\frac 1A
{\sum_{k=1}^{N} a_k \nu_k} =1$ into Eq. (\ref{eq:omegaaimdef}),
\begin{eqnarray}
  \Omega_{A,I,M}
  &= &
  \sum_{k=1}^{N}   \frac {a_k \omega_k}{A}
  \sum_{ 
    \langle\sum \nu_j a_j = A\rangle \atop \langle\sum \nu_j q_j = Q\rangle}
  \frac {\omega_k^{\nu_k-1}} {(\nu_k -1)!}
  \prod_{j\neq k} \frac{\omega_j^{\nu_j}} {\nu_j!}
  \nonumber\\*
  && {} \cdot \sum_{\{\nu_j\}} \xi\left(I,M|\{\nu_j\}\right).
  \label{eq:cl:Omega_tmp}
\end{eqnarray}
The entire system can be broken into two subsystems, a single particle with
isospin $I_k$ and projection $m_k$ and a remainder system with isospin $I'$ and
projection $M-m_k$, which are coupled with the appropriate Clebsch-Gordan
coefficients.  All possible values for the total isospin of the remainder
system have to be summed over,
\begin{eqnarray}
  \sum_{\{\nu_j\}} \xi\left(I,M|\{\nu_j\}\right)
  &=&
  \sum_{I'=|M-m_k|}^{I_k+I}  \sum_{\{\nu'_j\}}  \xi\left(I',M|\{\nu'_j\}\right)
  \nonumber\\*
  && {} \cdot 
  \langle I_k m_k; I', M-m_k| IM \rangle ^2
  .
\end{eqnarray}
With this modification, the summation indices in
\refEq{eq:cl:Omega_tmp} can be switched,
\begin{equation}
  \nu'_j = \left\{
    \begin{array}{ll}
      \nu_j     & , j\neq k \\
      \nu_j -1  & , j=k
    \end{array}
  \right.
\end{equation}
and the partition function written as recursion relation
\begin{eqnarray}
  \Omega_{A,I,M} 
  &=&
  \sum_{k=1}^N \frac{a_k\omega_k}{A}
  \sum_{I'=|I-I_k|}^{I+I_k}
  \Omega_{A-a_k,I',M-m_k}
  \nonumber \\*
  && {} \cdot
 \langle I_km_k; I',M-m_k | IM\rangle^2 
 .\label{eq:cl:Omega}
\end{eqnarray}

Since the partition function is the trace of an isoscalar, i.e. $e^{-H/T}$, it
will not depend on the isospin projection $M$. Hence, \refEq{eq:cl:Omega} can
be further simplified by summing the RHS over all isospin projections $m_k$ of
an isospin multiplet, 
\begin{equation}
  \Omega_{A,I}=\sum_{k'}\frac{a_k\omega_k}{A}\sum_{I^\prime=|I-I_k|}^{I+I_k}
  \Omega_{A-a_k,I^\prime},
\end{equation}
where the sum over $k'$ includes iso-multiplets, not individual
particles species.

\subsection{Degenerate Systems}
\label{subsec:partfunctheory_degen}

In a degenerate system several particles might occupy the same quantum state,
therefore, symmetrization of the wave function has to be accounted for.  For
the purpose of this paper we will restrict ourselves to studying Bose-Einstein
particles.  States with multiple particles have to be added to \refEq{eq:cl:Z},
which only contains states that are occupied by zero or one particle,
\begin{equation}
  \label{eq:qm:Z}
  Z_{A,M} =
  \sum_{n=1}^\infty \sum_{k=1}^N \frac{a_k}{A} C^{(k)}_n Z_{A-na_k, M-nm_k},
\end{equation}
where the cycle diagram
\begin{eqnarray}
  \label{eq:cycle}
  C^{(k)}_n&=&\langle\tilde{\alpha}|e^{-H/T}|\alpha\rangle
  \\
\nonumber
&=& \sum_l g_l\exp(-n\epsilon^{(k)}_l/T).
\end{eqnarray}
Here, the state $|\alpha\rangle$ refers to an n-particle state of
distinguishable particles and $|\tilde{\alpha}\rangle$ is the cyclic
permutation of that state. The single-particle energy levels are
$\epsilon_l^{(k)}$ for particle type $k$.  A more rigorous derivation of
\refEq{eq:qm:Z} is given in \cite{Pratt:1999ns}.  \\

Since the partition function for conserved total isospin $\Omega_{A,I,M}$
is independent of the isospin projection $M$, as mentioned above, a simple
relation can be derived, $ Z_{A,M}= \sum_{I\geq M} \Omega_{A,I}$, which
in turn leads to
\begin{equation}
  \label{eq:qm:Omega}
  \Omega_{A,I}= Z_{A,M=I}-Z_{A,M=I+1}.
\end{equation}

A second method for calculating the pion partition function constraining total
isospin is obtained by evaluating the cycle diagram $C_n^{(k)}$
in \refEq{eq:cycle} for its isospin content. These new cycle diagrams are
defined as 
\begin{eqnarray}
  \zeta_{n,i} &\equiv&
  \sum_{\beta} \langle\tilde{\beta},n,i| e^{-H/T} |\beta,n,i\rangle,
\end{eqnarray}
where the sum over $\beta$ represents a sum over all states with fixed particle
number $n$ and isospin $i$. The particles are assumed to be distinguishable and
$\tilde{\beta}$ represents a cyclic permutation of particles.  The partition
function for the pions in term of this new cycle diagram is then
\begin{equation}
  \label{eq:qm:Omega2}
  \Omega_{A,I} = \frac{1}{A} \sum_{n=1}^{A} \sum_{i=0}^{n}
  \sum_{I'=|I-i|}^{I+i} \zeta_{n,i} \Omega_{A-n,I^\prime},
\end{equation}
where the new cycle diagrams $\zeta$ are yet to be determined. After obtaining
a recursion relation for these functions  from  \refEq{eq:qm:Omega2} itself,
\begin{equation}
  \zeta_{A,I}= A \Omega_{A,I} - \sum_{n=1}^{A-1} \sum_{i=0}^{n}
  \sum_{I'=|I-i|}^{I+i} \zeta_{n,i} \Omega_{A-n,I^\prime},
\end{equation}
we find that these cycle diagrams follow a simple pattern by considering a
one-level system where $\Omega_{A,I}$ is easily calculated.
\begin{eqnarray}
  \zeta_{n,i}&=&\left\{
    \begin{array}{rl}
      C_n,&i=n\\
      -C_n,&i=n-1\\
      C_n,&i=0\\
      0,&\text{otherwise}
    \end{array}\right.
\end{eqnarray}

Since resonances are more massive and have lower phase space occupations, the
probability of creating several resonances in the same state can be neglected
except in the limit of extremely high densities. Therefore, resonances might
be treated as 
independent, non-degenerate subsystems, for which a partition function can be
obtained through \refEq{eq:cl:Omega}. The partition functions of two
subsystems, $1$ and $2$, can then be convoluted to obtain that of the entire
system,
\begin{equation}
  \label{eq:Omega_convolute}
  \Omega_{A,I,M} = \sum_{A'=0}^{A}\sum_{I'=0}^{A'} \sum_{I''=|I-I'|}^{I+I'}
  \Omega_{A',I'}^{(1)} \Omega_{A-A',I''}^{(2)} .
\end{equation}
If there are more than two subsystems \refEq{eq:Omega_convolute} can be applied
successively, i.e., partition functions of any two subsystems are
convoluted first to obtain a new partition function, which is
then convoluted with the partition function of another subsystem, and so on.

It should be pointed out that treating resonances as if they are in a different
system is not consistent with the indistinguishability of pions from resonances
and direct pions. For narrow resonances, one could uniquely identify the pions
by constructing the invariant masses of the constituents. However, for broad
resonances, one can not confidently identify the resonances. The criteria for
resonances being narrow are identical to the criteria that they have
sufficiently long lifetime to decay outside the collision region. For all the
calculations considered in this paper, it is assumed that the resonances are
separable. This assumption is excellent for pions from $\eta$ mesons,  good
for pions from $\omega$ resonances and questionable for
pions from $\rho$ decays. Including the effects of symmetrizing the resonant
and non-resonant pions remains an open question.

\section{Multiplicity Distributions and Isospin Fluctuations}
\label{sec:multdisttheory}

\subsection{Non-degenerate Systems}
\label{subsec:multdisttheory_classical}
The multiplicity distribution can be calculated from a ratio of partition
functions, where the numerator includes an extra constraint,
\begin{equation}
  \label{eq:cl:P_M:def}
  P_{A,M}(n_j) =  \frac{Z_{A,M,n_j}}{Z_{A,M}}.
\end{equation}
Here the numerator represents a canonical ensemble with the appropriate
conservation laws containing $n_j$ particles of type $j$. This additional
constraint can be regarded as a ``charge'' and, therefore, is added to the
indices of the partition function. The partition function in the numerator can
be rewritten with the aid of \refEq{eq:cl:Z},
\begin{equation}
  \label{eq:cl:P_M_tmp}
  Z_{A,M,n_j} = 
  \sum_{k=1}^N \frac{a_k \omega_k}{A} Z_{A-a_k, M-m_k, n_j-d_{k,j}} ,
\end{equation}
where the feed-down factor $d_{k,j}$ indicates that a particle of type $k$
decays into $d_{k,j}$ particles of type $j$.  This approach, however, will not
work for a resonance that decays via more than one decay channel. In such a
case, a pseudo-particle is included for each decay branch with its 
degeneracy $g_i$ scaled by the corresponding branching ratio.

It will prove more convenient to write equations in terms of the product of
partition function and multiplicity distribution
\begin{equation}
  W_{A,M}(n_j) \equiv  Z_{A,M} P_{A,M}(n_j)
\end{equation}
instead of the multiplicity distribution itself. For a non-degenerate system
conserving only additive charges, the multiplicity distribution can be
obtained from
\begin{equation}
  \label{eq:cl:W_M}
  W_{A,M}(n_j) = \sum_{k=1}^N \frac{a_k \omega_k}{A}
  W_{A-a_k,M-m_k}(n_j-d_{k,j}) .
\end{equation}

The occupation number for  particle type $j$  including feed-downs from
resonance decays is determined by multiplying  the
occupation number of all particles, given in \cite{Pratt:1999ht}, by the
feed-down-factor and summing over all resonances,
\begin{equation}
  \langle n_j \rangle =
  \frac 1{Z_{A,M}} 
  \sum_{k=1}^N  d_{k,j} \omega_k Z_{A-a_k, M- m_k}.
\end{equation}
The second moment of the distribution will be needed for calculating 
isospin fluctuations as in Sec.~\ref{subsec:multdisttheory_isofluc}.
\begin{eqnarray}
  \label{eq:classicalmoments}
  \langle n_j n_{j'} \rangle 
  &=&
  \sum_{k, k'=1}^N  \frac {d_{k,j} d_{k',j'}}{Z_{A,M}}   \bigr\{ 
  \delta_{k,k'} \omega_k  Z_{A-a_k, M- m_k} 
  \nonumber\\*
  && {} + \omega_k \omega_{k'} Z_{A-a_k -a_{k'}, M- m_k -m_{k'}} \bigr\}
  .
\end{eqnarray}

The multiplicity distribution incorporating conserved total isospin can be
derived in a similar manner as was employed for \refEq{eq:cl:Omega}.
\begin{eqnarray}
  W_{A,I,M}(n_j) &=&
  \sum_{k=1}^{N} \frac{a_k \omega_k} {A}
  \sum_{I'} \langle I_km_k; I',M-m_k | IM\rangle^2 
  \nonumber\\*
  && {} \cdot
  \label{eq:cl:P_I}
  W_{A-a_k,I',M-m_k}(n_j-d_{k,j}).
\end{eqnarray}

\subsection{Degenerate Systems}
\label{subsec:multdisttheory_degen}

When considering only quantum numbers corresponding to additive
charges, the multiplicity distribution is obtained through Eqs. (\ref{eq:qm:Z})
and (\ref{eq:cl:P_M:def}) and we find
\begin{equation}
  \label{eq:qm:W_M}
  W_{A,M}(n_j) =   \sum_{k=1}^{N} \frac{a_k}{A}
  \sum_{l=1}^{\infty} C^{(k)}_l W_{A-la_j,M-lm_j}(n_j-ld_{k,j}),
\end{equation}
where $C_l$ is the cycle diagram define in Eq. (\ref{eq:cycle}). Calculating
the two-point function and the four-point functions allows one to obtain the
first two moments of the multiplicity distribution.  The 2-point function is
\begin{equation}
  \label{eq:2-point}
  \langle a_i^\dagger a_j\rangle = \frac{\delta_{ij}}{Z_{A,M}}
  \sum_n \exp(-n\epsilon_i/T) Z_{A-na_{i},M-n m_{i}}
\end{equation}
Summing over all particle types and states and multiplying by feed-down factors
results in an expression for the occupation numbers,
\begin{equation}
  \langle n_j \rangle
  = \frac 1{Z_{A,M}} 
  \sum_{k=1}^N  d_{k,j} \sum_{l=1}^{\infty} C^{(k)}_l Z_{A-la_k, M- lm_k}.
\end{equation}
Similarly, the 4-point function 
\begin{eqnarray}
  \label{eq:4-point}
  \langle a_i^\dagger a_j^\dagger a_k a_l\rangle 
  &=&
  \frac{\delta_{il}\delta_{jk} + \delta_{ik}\delta_{jl}}{Z_{A,M}}
  \nonumber \\*
  && {} \cdot
  \sum_{n_i,n_j} 
  \exp(-n_i\epsilon_i/T)
  \exp(-n_j\epsilon_j/T) 
  \nonumber \\*
  && {} \cdot
  Z_{A-n_i a_i-n_j a_j, M-n_i m_i-n_j m_j}
\end{eqnarray}
serves to obtain second moments of the multiplicity distribution
\begin{eqnarray}
  \label{eq:cross_variance}
  \lefteqn{\langle n_j n_{j'} \rangle =
    \frac{\delta_{j,j'}}{Z_{A,M}} \sum_{l,l'} C_{l+l'} 
    Z_{A-(l+l') a_j, M- (l+l') m_j}
  }  \nonumber\\*
  && {} + \frac 1{Z_{A,M}} 
  \biggr\{ \sum_{l} 
  \delta_{j,j'} C_l  Z_{A-la_k, M- lm_k}
  \nonumber\\* 
  && {} + \sum_{l,l'} C_l  C_{l'}
  Z_{A-l a_k -l' a_{k'}, M- l m_k -l' m_{k'}} \biggr\}
  .
\end{eqnarray}
Neglecting the terms with $C_\ell$ where $\ell>1$, one returns to the
non-degenerate result, \refEq{eq:classicalmoments}.
\\

Calculating the multiplicity distribution for degenerate systems with the
constraint of total isospin conservation becomes difficult because
the analog of the cycle diagram, $C_n^{(k)}$ in \refEq{eq:qm:W_M}, needs
to be analyzed for isospin $i$ and charge $n_j$. Such a cycle
diagram with $a$ particles and isospin projection $m$ can be written as
\begin{eqnarray}
  \chi_{a,i,m}(n_j)&\equiv&
  \sum_{\alpha,\beta}
  \langle\tilde{\alpha},a,i,m| e^{-H/T}
  |\beta,a,n_j,m\rangle\\
  \nonumber
  &&\hspace*{60pt}\cdot\langle\beta,a,n_j,m
  |\alpha,a,i,m\rangle,
\end{eqnarray}
where the sums over $\alpha$ and $\beta$ correspond to sums over all states
with fixed $(a,i,m)$ and $(a,n_j,m)$, respectively, and $\tilde{\alpha}$
represents a cyclic permutation of particles, which  are assumed to be
distinguishable. 
 
The multiplicity distribution can be calculated in terms of these cycle diagrams
\begin{eqnarray}
\label{eq:qm:wrecursion}
W_{A,I,M}(n_j) &=& \frac{1}{A}\sum_{a,i,m,n'_j,I'}
\chi_{a,i,m}(n'_j)\\
\nonumber
&&
\hspace*{-40pt}\cdot W_{A-a,I^\prime,M-m}(n_j-n'_j)
\langle I^\prime,M-m;i,m|I,M\rangle^2
\end{eqnarray}
To derive the cycle diagrams $\chi$, consider a single-level system, for which
the probability distribution $W^{(1)}$ is calculated in Appendix
\ref{appendix_zwonelevel}. It follows from \refEq{eq:qm:wrecursion} that
\begin{eqnarray}
  \chi^{(1)}_{A,I,M}(n_j)&=& A\, W^{(1)}_{A,I,M}(n_j)
  -\sum_{a< A,i,m,n_j,I^\prime}\chi^{(1)}_{a,i,m}(n_j)
  \nonumber \\* \nonumber &&
  \hspace*{-40pt}\cdot W^{(1)}_{A-a,I^\prime,M-m}(n_j-n_j)
  \, \langle I^\prime,M-m;i,m|I,M\rangle^2 \\
  \label{eq:chi^1}
\end{eqnarray}
The function $\chi$ which accounts for all levels can be easily generated from
$\chi^{(1)}$,
\begin{equation}
\chi_{a,i,m,n_k}=\chi^{(1)}_{a,i,m,n_k}\, \sum_\ell g_\ell e^{-aE_\ell/T},
\end{equation}
where $\ell$ indicates the single-particle energy levels with energy $E_\ell$.

If different particles, like resonances, are to be included in
the ensemble, one can either insert a sum over species into 
\refEq{eq:qm:wrecursion}; or calculate $W$ separately for each
species and convolute them to find $W$ for the entire system,
\begin{eqnarray}
  \lefteqn{W_{A,I,M}(n_j)  = \sum_{A',I',M',n'_j,I''} 
  W_{A',I',M'}^{(1)}(n'_j) }
  \\ \nonumber
  && {} \cdot
  W_{A-A',I'',M-M'}^{(2)}(n_j-n'_j)   \langle I'M'; I'',M-M' | IM\rangle^2.
  \label{eq:W_convolute}
\end{eqnarray}
In our calculations, $W$ was calculated separately for resonances neglecting
symmetrization and  then convoluted with $W$ calculated for pions with proper
symmetrization.

\subsection{Isospin Fluctuations}
\label{subsec:multdisttheory_isofluc}

Given some system with quantum states $\alpha$, pion isospin fluctuations can
be defined as  
\begin{equation}
  \label{eq:G_def}
  G^2= \sum_{\alpha} 
  \left\langle \alpha \left| (N_+ + N_- - 2N_0)^2 \right| \alpha \right\rangle,
\end{equation}
where $N_+$, $N_-$ and $N_0$ are the number operators of the respective pions.
These isospin fluctuations could be computed through multiplicity distributions
or with the expressions for densities and higher moments that were given above.
However, when total isospin should be conserved as well, 
multiplicity distribution calculations are slow and expressions for densities
and higher moments are difficult to derive.  Instead, we will
derive the isospin fluctuation for a system in an isosinglet in terms of
isospin projection states.

The operator in \refEq{eq:G_def} is a product of two rank-2 spherical
tensors components
\begin{equation}
  (N_+ + N_- - 2N_0)^2 = 6 T_{20}T_{20} ,
\end{equation}
where
\setlength{\arraycolsep}{1.5pt}
\begin{subequations}
  \begin{eqnarray}
    T_{20} &=& \sum_i \frac1{\sqrt{6}} \left(
      \pi_{+,i}^\dagger \pi_{+,i} + \pi_{-,i}^\dagger \pi_{-,i} 
      - 2  \pi_{0,i}^\dagger \pi_{0,i}
    \right), \quad\quad
    \\
    T_{2\pm 1} &=& \sum_i \frac1{\sqrt{2}} 
    \left(\pi_{0,i}^\dagger \pi_{\mp,i} - 
      \pi_{\pm,i}^\dagger\pi_{0,i} \right),
    \\
    T_{2\pm 2} &=& \sum_i \pi_{\pm,i}^\dagger \pi_{\mp,i}.
\end{eqnarray}
\end{subequations}
A product of spherical tensors can be decomposed into other spherical tensor
components 
\begin{equation}
  T_{20} T_{20}
  =  \sum_{J,M} \langle 20; 20 | JM \rangle A_{JM}.
\end{equation}
By the Wigner-Eckart theorem only $A_{00}$ contributes when contracted between 
isosinglet states, and because $A_{00}$ is an isoscalar we can write
\begin{equation}
  \langle G^2\rangle = \frac 6{\sqrt{5}} \sum_{I=0} 
  \langle A_{00} \rangle
  = \frac 6{\sqrt{5}} \left(\sum_{M=0}-\sum_{M=1}\right) \langle M | A_{00} | M \rangle.
\end{equation}
After some algebra we obtain
\begin{equation}
  A_{00} = \sum_M \langle 2M; 2,-M | 00 \rangle T_{2M}T_{2,-M}
  = A_{00}' + A_{00}^{\text{QM}} ,
\end{equation}
where
\begin{equation}
  A_{00}' = \frac 1{\sqrt{5}}
  \left[\frac 32 N_+ + \frac 32 N_- +  N_0 + 
    \frac 16 \left(N_+ + N_- - 2N_0\right)^2 \right]
\end{equation}
and
\begin{eqnarray}
\label{eq:a00qmdef}
  A_{00}^{\text{QM}} &=& \frac 1{\sqrt{5}}
  \sum_{i,j} \left(  
    2 \pi_{+,i}^\dagger \pi_{-,j}^\dagger \pi_{+,j} \pi_{-,i}  +
    \pi_{+,i}^\dagger \pi_{0,j}^\dagger \pi_{+,j} \pi_{0,i}  
  \right. \nonumber\\*
  && {} + \left. \pi_{0,i}^\dagger \pi_{-,j}^\dagger \pi_{0,j} \pi_{-,i}  
  \right)
  .
\end{eqnarray}
The expectation of $A_{00}^{\text{QM}}$ is non-zero when particles of different
charges are in the same quantum state, or when two differently charged pions
are produced into two different states with a quantum correlation arising from
a resonance decay.  The contribution to $A_{00}^{\text{QM}}$ from the
degenerate nature of the pion states can be determined via \refEq{eq:4-point},
\begin{eqnarray}
  \lefteqn{
    \langle\pi_{k,i}^\dagger \pi_{k',j}^\dagger \pi_{k,j} \pi_{k',i}\rangle
    = }
  \nonumber \\* &&
  \frac {1}{Z_{A,M}} \sum_{l,l'} C^{(\pi)}_{l+l'} 
  Z_{A-l a_k -l' a_{k'}, M- l m_k -l' m_{k'}}.
\end{eqnarray}
This contribution can be ignored in the non-degenerate limit, where
occupation numbers are small.

Contributions to $A_{00}^{\text{QM}}$ from the coherent correlation between
pions 
from resonant decays can be found by expressing the resonances in terms of pion
creation operators. For example, the $\rho^+$ meson, which is a member of an
isotriplet, can be considered as one pion in an $s$ wave and a second pion in a
$p$ state. Referring to these two states as $i$ and $j$,
\begin{equation}
  |\rho^+\rangle=\frac{1}{\sqrt{2}}\left(\pi_{+,i}^\dagger \pi_{0,j}^\dagger
    -\pi_{+,j}^\dagger \pi_{0,i}^\dagger\right)|0\rangle.
\end{equation}
How the states $i$ and $j$ are chosen is irrelevant since they are summed over
in \refEq{eq:a00qmdef}, but the coherent mixture of the two permutations,
which is necessary for the $\rho^+$ to be a member of an isotriplet, results in
a non-zero contribution to $A_{00}^{\text{QM}}$,
\begin{equation}
  \langle A_{00}^{\text{QM}}\rangle
  = -\frac{1}{\sqrt{5}}\left(2N_{\rho^0}+N_{\rho^+}+N_{\rho^-}\right).
\end{equation}
The $\omega$ and $\eta$ mesons are isosinglets and can be treated accordingly.
For instance,
\begin{equation}
  |\eta\rangle=\frac{1}{\sqrt{6}}\sum_{i,j,k}\epsilon_{ijk}
  \pi_{+,i}^\dagger \pi_{-,j}^\dagger \pi_{0,k}^\dagger |0\rangle,
\end{equation}
which adds another term to $A_{00}^{\text{QM}}$,
\begin{equation}
  \langle A_{00}^{\text{QM}}\rangle=
  -\frac{4}{\sqrt{5}}\left(N_\omega+N_\eta\right).
\end{equation}

\section{Results}
\label{sec:results}

In heavy ion reactions, and perhaps in $pp$ reactions, pions reinteract with
other pions in their neighborhood, or domain, and might be expected to sample a
large portion of the the available phase space. As isospin should be conserved
in each domain, it seems reasonable to explore distributions for a few dozen
pions rather than creating an ensemble of a few thousand pions, which could be
treated as a grand canonical ensemble \cite{gavin}. One of our study's goals
is 
to understand how many pions are required for conservation constraints to
become irrelevant.

In the following, symmetrization and resonances are first ignored in order to
focus on the effects of conserving isospin, subsequently, the
effects of symmetrization and resonances are illustrated by
considering a simple example.

\subsection{Total Isospin Conservation}
\label{subsec:results_classical}

\begin{figure}
  \includegraphics[width=0.45\textwidth]{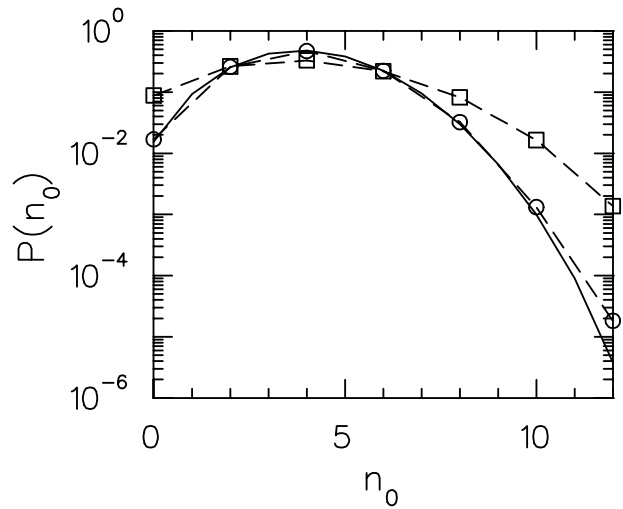}
  \caption{\label{fig:isodist_isoconserve} 
    The probability of observing $n_0$ neutral pions is shown for a system of
    12 pions. Symmetrization and resonances are neglected. A random
    distribution, multiplied by 2 for comparison, is represented by a solid
    line and is used as a benchmark. A distribution resulting from isoscalar
    pairs of pions (squares) is significantly broader, whereas a distribution
    including all isosinglets (circles) approaches the random distribution.}
\end{figure}

When quantum degeneracy and resonances are ignored, isospin
distributions are independent of energy levels or
temperature. Therefore, the results presented in the following are generic to
any system 
where only pions are considered and the phase space occupation
numbers are small.
A random distribution ignoring isospin conservation, i.e. a mixed-event
construction, will serve as a benchmark.
\begin{equation}
  P_{\rm random}(n_0)=\left(\frac{1}{3}\right)^N\sum_{n_++n_-+n_0=N}
  \frac{N!}{n_+!n_-!n_0!}
\end{equation}
Unlike distributions that conserve isospin, this distribution allows both even
and odd numbers of neutral pions and is, therefore, scaled by a factor of two
to compare the width with that of the other distributions.
Secondly, when pion creation is constrained to isoscalar pairs, as
in \cite{Pratt:1994tu}, the distribution can be considered as a binomial
distribution of pairs where one third of the time the pair is comprised of two
neutral pions and two thirds of the time the pair is comprised of a positive
and negative pion.
\begin{eqnarray}
  P_{\rm pairwise}(n_0)&=&\left(\frac{1}{3}\right)^{n_0/2}
  \left(\frac{2}{3}\right)^{(N-n_0)/2}\\
  \nonumber
  &&\hspace*{40pt}\cdot\frac{(N/2)!}{(n_0/2)!(N/2-n_0/2)!}
\end{eqnarray}
This pairwise distribution is broader  than the
random distribution by a factor of $\sqrt{2}$, as can be seen in
Fig.~\ref{fig:isodist_isoconserve}.

Finally, the isospin distribution for non-degenerate particles is calculated
with the methods of Sec.~\ref{subsec:multdisttheory_classical} with all 12-pion
isosinglet states being considered.  The constraint of exact isospin
conservation only modestly broadens the distribution relative to the random
distribution as shown in Fig.~\ref{fig:isodist_isoconserve}.

\begin{figure}
  \includegraphics[width=0.45\textwidth]{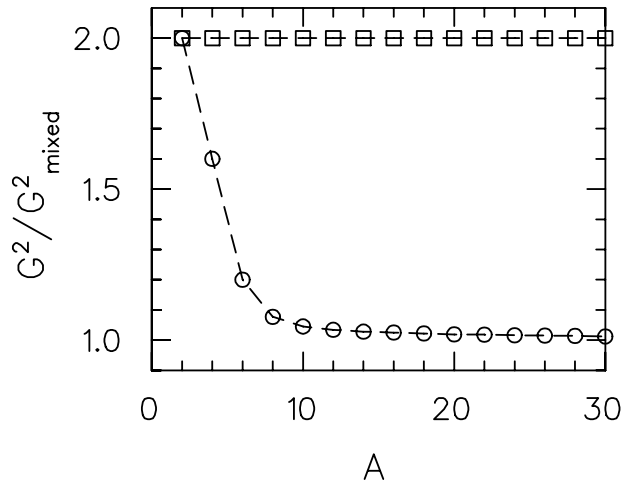}
  \caption{\label{fig:classmoments_vs_A} 
    The fluctuation $G^2$ for non-degenerate systems as a function of 
    system size for an ensemble of isoscalar pairs (squares) and one including
    all isosinglet states (circles). The fluctuations have been
    normalized by the fluctuation for a random system. The fluctuations for the
    pairwise distribution are twice the fluctuations for a random distribution,
    whereas including all isosinglet states relaxes the constraint and results 
    in the same width as the random distributions for large systems.}
\end{figure}

These findings are underscored by comparing the isospin fluctuations as a
function of total pion number, as shown in Fig.~\ref{fig:classmoments_vs_A}.
When pion emission is constrained to isoscalar pairs, fluctuations are twice as
large as compared to the random case for all system sizes. When considering all
$N$-pion isosinglets fluctuations are larger by a factor which falls from two
to unity as $N$ approaches infinity.

\subsection{Including Symmetrization and Resonances}
\label{subsec:results_be_resonances}

To illustrate the effects of Bose-Einstein symmetrization, total
isospin conservation, and resonance decays, an assumption must be made about
the available single-particle energy levels that are summed over in the
single-particle partition function.  Assuming the model system is
confined to a cube of volume $V$, the energy states are obtained
\begin{equation}
  \epsilon_{n,m,l}
  =
  \sqrt{ \frac{\pi^2}{R^2} \left( n^2 + m^2 + l^2 \right) + M^2} ,
\end{equation}
where $R= V^{1/3}$, $M$ is the mass of the particle, and 
$n,m,l$ are chosen to be half-integers.
The choice of half integers, instead of the more usual integers,  deemphasizes
zero-point surface energy effects and seems more physical if the confinement
to the volume does not arise from an infinite potential well. This becomes
important when systems are confined to a small volume. 

Isospin distributions were calculated for a system of 24 pions at two
densities, 0.3~fm$^{-3}$, which is well above breakup densities for hadronic
collisions and 0.1~fm$^{-3}$. The temperature was chosen to be 125 MeV.  
The Bose-Einstein nature of pions was taken into account when  calculating
multiplicities according to the formalism from
Sec.~\ref{subsec:multdisttheory_degen}. The multiplicity distributions
were then convoluted with those of  three resonant states, the
isotriplet $\rho$ mesons and the isosinglet $\omega$ and $\eta$ mesons,
which were treated as non-degenerate systems according to
Sec.~\ref{subsec:multdisttheory_classical}. 
In principle, strange mesons and baryonic resonances can be incorporated as
well, but the calculation would be significantly lengthened by the inclusion
of extra indices. Prospects for such calculations are discussed in the
conclusions.

\begin{figure}[htb]
  \includegraphics[width=0.45\textwidth]{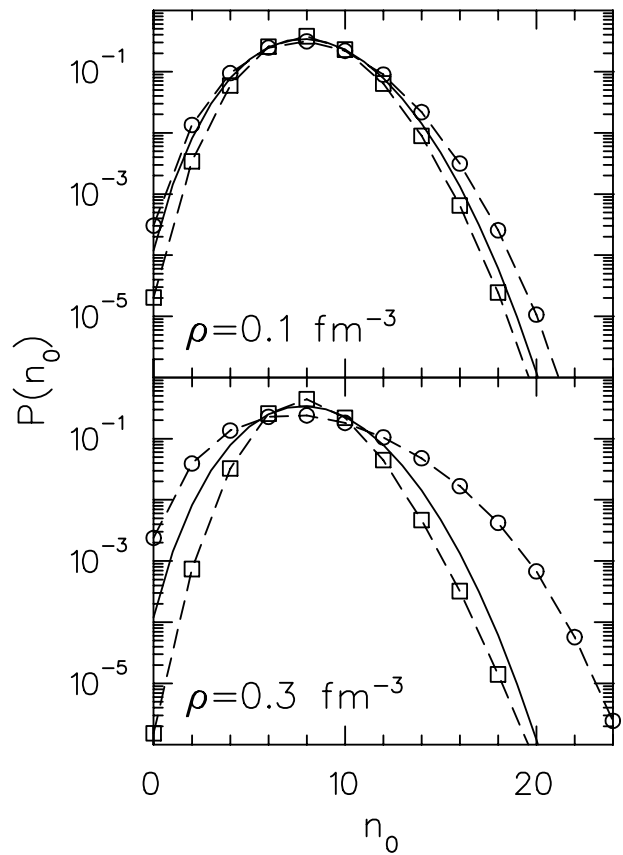}
  \caption{\label{fig:isodist}
    The probability for producing $n_0$ neutral pions in a system of $A=24$
    pions at $T=125$ MeV is shown for ensembles restricted to an overall
    isosinglet and with symmetrization included. Calculations with resonances
    (squares) and without (circles) are displayed. Although symmetrization
    broadens the distribution relative to the random distribution (solid
    line), the inclusion of resonances results in a narrower distribution
    as compared to the random distribution.} 
\end{figure}

\begin{figure}[htb]
  \includegraphics[width=0.45\textwidth]{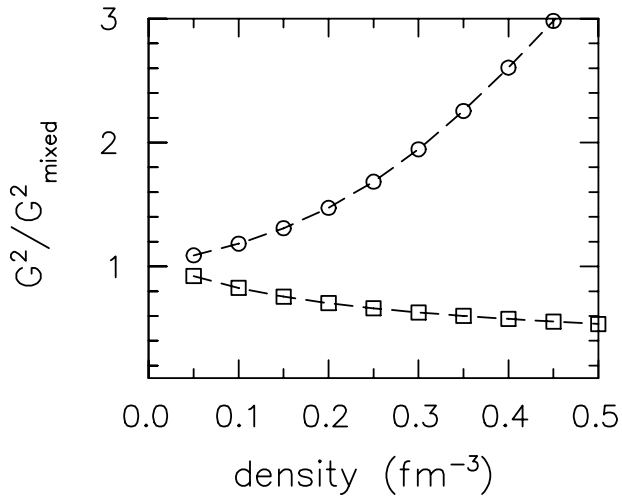}
  \caption{\label{fig:moments_vs_dens}
    Isospin fluctuations, or squared width of the isospin distributions,
    as a function of density scaled by the width of a random
    distribution. Calculations were performed for 24 pions  restricted to
    isosinglet states at a temperature of 125 MeV. 
    At high density, the distributions are broadened by including symmetrization
    and narrowed by including resonances produces.}
\end{figure}

As expected, the isospin distributions for symmetrized pions in an
isosinglet are broader than the random distribution when resonances are
neglected, as shown in Fig.~\ref{fig:isodist}. 
This broadening is especially pronounced at high density. 
However, the inclusion of resonances more than compensates for the
symmetrization effects and results in distributions that are narrower than
the random distribution.
Figure \ref{fig:moments_vs_dens} displays fluctuations as a function of
density for the 24-pion system. The dramatic broadening
induced by symmetrization  at high density is counteracted by a remarkable
narrowing when resonant states are considered.

The conditions for pions to prefer forming a resonance are similar to the
conditions for symmetrization to be important, i.e., a high phase space
density. At low temperature, where  the mass penalty for resonant formation
would play a larger role, resonant effects would become relatively less 
important than symmetrization. 
However, for the temperature of 125 MeV considered here, the resonant effects
overwhelm the effects of symmetrization at all densities.
 
Finally, it should be noted that the isospin fluctuation $G^2$ was calculated
both from the distributions themselves and from the methods described in
Sec.~\ref{subsec:multdisttheory_isofluc}. Although the two sets of numerical
calculations  have little in common aside from the functions used to generate
the single-particle levels, the two sets of moments agreed with each other
within the numerical accuracy of the computer.

\section{Conclusions}
\label{sec:conclusions}

For the first time, expressions were obtained for multiplicity distributions
and isospin fluctuations for a canonical ensemble, in which total isospin as
well as additive quantum numbers are exactly conserved.  The formalism has been
extended to include both pions and resonances and can account for Bose-Einstein
symmetrization of the pion wavefunction. Numerical calculations were then
performed to study the effects of total isospin conservation, quantum
symmetrization, and resonance decays on the width of the multiplicity
distribution, which can be squared to obtain the isospin fluctuations. Direct
expressions for the isospin fluctuations permit a much quicker calculation than
those for the multiplicity distributions.

It was found that conservation of total isospin and its projection has little
effect on the width of the multiplicity distributions, when the systems are
larger than a dozen particles. It should be noted that conserving only the
projection, not the total isospin, would result in distributions where the
average number of neutral pions would not equal one third of the total. At high
phase space densities, including Bose-Einstein symmetrization leads to a
multiplicity distribution that is much broader than a random distribution.
However, addition of resonances more than compensates for this broadening and
narrows the multiplicity distribution below the width of the random
distribution.  Both effects are small when the phase space density is below
0.1~fm$^{-3}$.

The widths of the multiplicity distributions are largely dominated by the
behavior of the tails, thus making it imperative to perform exact
calculations. These calculations were made possible by using recursion
relations that circumvent summing over the immense number of partitions in the
partition functions. 

The calculations in this paper were largely schematic and included only three
meson resonance, $\rho$, $\eta$, and $\omega$.  For a more realistic
calculation more resonances like strange mesons and baryonic resonances need to
be added along with strangeness and baryon number conservation.  The formalism
presented in this paper scales linearly with the number of particle species if
the resonances were to be included without additional conserved charges.
Theoretically, any amount of quantum numbers can be conserved as long as they
commute with the isospin operator. Practically, every new index, which has to
be added to the partition function and multiplicity distributions, to conserve
another charge increments the number of loops by one. The increase in runtime
has little consequence for calculations of partition functions and direct
calculations of isospin fluctuations, which are virtually instantaneous,
whereas any additional index would significantly slow down calculations of
multiplicity distributions, which take on the order of ten minutes.

The aforementioned caveats are not expected to become major obstacles in the
application of the presented formalism. Although the particle multiplicities
are high, in the thousands, in possible physical applications like relativistic
heavy ion collisions, the local nature of charge conservation would limit the
number of particles considered at any given time.  The system under
consideration would have to be broken into domains, in which the charges are
conserved locally, then calculations proceed one domain at a time. Each domain
would have a relatively small number of particles with which to cope.

Before tackling the more numerically challenging problem of including
strangeness and baryon number, one should consider the limitations of any
comparison with experiment. Most importantly, it difficult to count neutral
pions as each neutral pion decays into two photons. Furthermore, one should
consider the ability to identify neutrons and kaons, especially those kaons
which then decay into pions. Given the inherent complexity of any such
measurement, we felt that it was proper to stop short of performing more
complicated calculations without a commensurate consideration of the
measurement. Nonetheless, several valuable lessons were gained from the
schematic calculations presented here.

Another possible application of partition functions with exact quantum number
and isospin conservation, one that has not been explored in this paper, are
Monte Carlo algorithms for particle generation.  Recently, there has been much
interest in modeling relativistic heavy ion collisions with hybrid models, in
which early, dense stages of the collision are described by a hydrodynamical
model before switching to a hadronic cascade to simulate the freeze-out stage
\cite{Bass:2000ib}.  The change of degrees of freedom at the interface between
the two models from the energy-momentum tensor to hadrons is generally modeled
by a grand-canonical ensemble, which conserves charges only in the average over
many events. However, event-by-event charge conservation is essential to
calculating observables like fluctuations and balance functions, which have
been proposed as possible signal for the quark-gluon-plasma \cite{Bass:2000az}.

\begin{acknowledgments}
  This work was supported by the National Science Foundation, Grant No.
  PHY-00-70818.
\end{acknowledgments}

\appendix

\section{Single-level Multiplicity Distribution Conserving Total Isospin} 
\label{appendix_zwonelevel}

By definition, the distribution $W^{(1)}_{a,i,m}(n_j)$, needed in
\refEq{eq:chi^1}, can be written as the product of the partition function and
the multiplicity distribution,
\begin{equation}
  W^{(1)}_{a,i,m,n_j}= \omega_{a,i,m}^{(1)} p_{a,i,m}^{(1)} (n_j),
\end{equation}
where the partition function for $a$ particles in a single level with
energy $E$ is
\begin{equation}
  \label{eq:qm:omega^1}
  \omega_{a,i,m}^{(1)} =
  \left\{
    \begin{array}{ll}
      \exp(-a E/T)  & , \text{if}\ a+i\ \text{even}\\
      0 & , \text{if}\ a+i\ \text{odd}
    \end{array}
  \right.
\end{equation}
$p_{a,i,m}^{(1)} (n_j)$ is the probability of observing $n_j$ pions in a single-state system
that contains a total of $a$ pions with total isospin $i$ and projection $m$.
This probability distribution has to be calculated  for only one type of
pions because the pion occupation numbers are related
through 
\begin{eqnarray}
  m &=& n_+ - n_-,\\
  a &=& n_0 + n_+ + n_-.
\end{eqnarray}
In the following, probability distributions  will be derived for positive
pions.

The isospin wave function of the system can be written in terms of eigenstates of the number operators
\begin{equation}
  \label{eq:aim_state}
  |a,i,m\rangle = \sum_{n_+=m}^{(a+m)/2}
  \alpha_{a,i,m,n_+} |n_0\rangle |n_+\rangle |n_-\rangle,
\end{equation}
where $ n_0 = a+m -2 n_+$ and $  n_- = n_+ -m$. The coefficients in
\refEq{eq:aim_state} are related to the probability distribution by
\begin{equation}
  p_{a,i,m}^{(1)} (n_+)= (\alpha_{a,i,m,n_+})^2.
\end{equation}
The isospin wave function $|a,a,a\rangle$ can only be constructed if  all pions
in the state are positive, i.e., $n_+= a$, therefore 
\begin{equation}
  \label{eq:alpha_aaan}
  \alpha_{a,a,a,n_+} = \left\{
  \begin{array}{ll}
    1 & , \text{if}\ n_+= a\\
    0 & , \text{otherwise}
  \end{array}
  \right.
\end{equation}
Leaving the pion number $a$ and isospin $i=a$ fixed, we can apply the isospin
lowering operator  
$I_- = \sqrt{2}(\pi_-^\dagger\pi_0 + \pi_0^\dagger\pi_+)$
to reach lower values of $m$,
\begin{equation}
  \label{eq:LoweringOpEffect}
  I_- |a,i,m\rangle = \sqrt{i(i+1)-m(m-1)}|a,i,m-1\rangle .
\end{equation}
The LHS of \refEq{eq:LoweringOpEffect} expands to
\begin{widetext}
\begin{eqnarray}
  \lefteqn{
    I_- \sum_{n_+}\alpha_{a,a,m,n_+} |a+m-2n_+\rangle |n_+\rangle |n_+-m\rangle
  }
  \nonumber\\* && 
  =
  \sqrt{2} \sum_{n_+=m}^{(a+m)/2} 
  \sqrt{a+m-2n_+} \sqrt{n_+ -m +1} \alpha_{a,a,m,n_+} 
  |a+(m-1)-2n_+\rangle |n_+\rangle |n_+-(m-1)\rangle 
  \nonumber\\* && {} +
  \sqrt{2} \sum_{n_+=m-1}^{(a+m)/2-1}
  \sqrt{a+(m-1)-2n_+} \sqrt{n_+ +1}  \alpha_{a,a,m,n_+ + 1} 
  |a+(m-1)-2n_+\rangle |n_+\rangle |n_+-(m-1)\rangle .
\end{eqnarray}
When the coefficients on the LHS  are match with those on the RHS of
\refEq{eq:LoweringOpEffect} a recursion relation is obtained,
\begin{eqnarray}
  \alpha_{a,a,m-1,n_+} =
  \sqrt{\frac{2} {i(i+1)-m(m-1)}} &\biggl\{&
  \sqrt{(a+m-2n_+)(n_+ -m +1)} \alpha_{a,a,m,n_+} 
  \nonumber\\* && 
  + \sqrt{(a+m-1-2n_+)(n_+ +1)}  \alpha_{a,a,m,n_+ + 1} 
  \biggl\}
  \label{eq:alpha_aam-1n}
\end{eqnarray}
\end{widetext}

So far, we have only found the coefficients $\alpha_{a,i,m,n_+}$ for
$a=i$. With the help of the isoscalar operator
\begin{equation}
  \label{eq:U_2_def}
  U_2 = 2\pi_+^\dagger\pi_-^\dagger - \pi_-^\dagger\pi_-^\dagger
\end{equation}
that creates two pions without altering the isospin,
higher values of $a$ can be reached,
\begin{equation}
  \label{eq:U_2}
  U_2 |a,i,m\rangle = N_{a,i} |a+2,i,m\rangle ,
\end{equation}
where the normalization constant is
\begin{equation}
  N_{a,i}= \sqrt{(a+2)(a+3)-i(i+1)}.
\end{equation}
Matching the coefficients on both sides of \refEq{eq:U_2} to each other leads
to 
\begin{widetext}
  \begin{equation}
    \alpha_{a+2,i,m,n_+} = \frac 1N \biggl\{
    2 \sqrt{n_+(n_+ - m)} \alpha_{a,i,m,n_+ -1} 
    -    \sqrt{(a+m-1-2n_+)(a+m-2-2n_+)}  \alpha_{a,i,m,n_+} 
    \biggl\} 
    .
   \label{eq:alpha_a_2imn}
 \end{equation}
\end{widetext}

In case, $a+i$ is odd, no combination of pions yields the isospin wave
function $|a,i,m\rangle$, as is evident from \refEq{eq:qm:omega^1}. Therefore
\begin{equation}
  \label{eq:alpha_a+i_odd}
  \alpha_{a,i,m,n_+}= 0\quad ,\ \text{if}\ a+i\ \text{odd}.
\end{equation}

Equations
(\ref{eq:alpha_aaan}), 
(\ref{eq:alpha_aam-1n}),
(\ref{eq:alpha_a_2imn}), and
(\ref{eq:alpha_a+i_odd}) 
completely determine the coefficients $\alpha_{a,i,m,n_+}$, which in turn
define the probability distributions $p^{(1)}_{a,i,m} (n_+)$ and, therefore,
the single-level partition functions $W_{a,i,m,n_j}^{(1)}$.


\end{document}